\documentclass[11pt]{amsart}
\usepackage{geometry}
\usepackage{amsmath}
\usepackage{amssymb}
\usepackage{amsfonts}
\usepackage{xcolor}
\usepackage{amsthm}
\usepackage{todonotes}
\usepackage{enumerate}
\usepackage{enumitem}
\usepackage{setspace}
\usepackage{mathtools}
\usepackage{graphicx}
\usepackage{soul}
\usepackage{biblatex}
\usepackage{hyperref}
\usepackage{xfrac}
\usepackage{nicefrac}
\usepackage{tikz-network}
\usepackage[justification=centering]{caption}
\usepackage{subcaption}
\usepackage{booktabs}
\usepackage{multirow}

\newtheorem{theorem}{Theorem}
\newtheorem{lemma}{Lemma}
\newtheorem{proposition}{Proposition} 
\newtheorem{corollary}{Corollary}

\theoremstyle{definition}
\newtheorem{definition}{Definition}

\newtheorem{example}{Example}

\AtBeginEnvironment{example}{%
  \pushQED{\qed}%
}
\AtEndEnvironment{example}{\popQED\endexample}

\newcommand{\bbR}{\mathbb{R}}

\newcommand{\calL}{\mathcal{L}}

\providecommand{\keywords}[1]
{
  \small	
  \textbf{\textit{Keywords---}} #1
}

\begin{document}

\title{Diffusion in dynamic networks with time-varying inputs to allocate responsibility}

\author[R. van den Ende]{Rosa van den Ende}

\address{Centre d'Économie de la Sorbonne, Université Paris 1 Panthéon-Sorbonne, Center for Mathematical Economics, Universität Bielefeld}
\email{rosa.ende@gmail.com}

\author[D. Laplace Mermoud]{Dylan Laplace Mermoud} 

\address{Unit{\'e} de Math{\'e}matiques Appliqu{\'e}es, ENSTA - Institut Polytechnique de Paris \& CEDRIC, Conservatoire National des Arts et M{\'e}tiers}
\email{dylan.laplace.mermoud@protonmail.com}

\setstretch{1.3}
\date{\today}

%\thanks{We wish to thank Mayeul Chavanne and Agnieszka Rusinowska for the helpful discussions with them. This work has received funding from the European Union’s Horizon 2020 research and innovation programme under the Marie Skłodowska-Curie grant agreement No 956107, “Economic Policy in Complex Environments (EPOC)”. This research was done while the second author benefited from the support of the FMJH Program PGMO, under project number 2023-0009, as well as from the ANR project HQI-ANR-22-PNCQ-0002. }


\begin{abstract}
Responsibility in complex networks extends beyond direct actions: players should also bear responsibility for the indirect effects within their supply chains or network. We introduce a novel framework to allocate responsibility for indirect environmental, social, and economic impacts across a dynamic network. Unlike static approaches, our framework accounts for the evolving structure of supply chains, financial systems, and other interconnected systems, where relationships change over time. We use the time-dependent Laplacian matrix to capture how responsibility propagates through the network, revealing a diffusion process that aligns with key axioms of fairness: linearity, efficiency, symmetry, and the independent player property. We show that approximating the responsibility measure preserves these properties, supporting the use of our framework as a rigorous method to allocate responsibility in real-world networks. 
\end{abstract}

\maketitle

\section{Introduction}

The growing complexity of the global economy, with interconnected supply chain or networks spanning multiple industries and countries, has created significant challenges. It has become increasingly difficult to determine what the indirect effects of one's actions or operations are, and subsequently, to hold the players responsible for these effects. Against this background, more and more regulations to stimulate transparency have been put in place such as the European Corporate Sustainability Reporting Directive (CSRD) \parencite{EU_CSRD} or the US SEC rules \parencite{SEC_Press_Release_2022-46}, that require certain firms to disclose not only their financial statements, but also their environmental, social and governance (ESG) impacts. These sustainability impacts of a firm concern the effects of its own operations as well as those of its buyers and suppliers. While these regulations are designed to ultimately hold companies responsible for their ESG impacts, this notion of responsibility is not yet properly quantified. Moreover, supply chains and financial or production networks are inherently dynamic: they evolve over time as firms adjust suppliers, enter or exit markets, or respond to economic and regulatory pressures. These structural changes affect how responsibility propagates through the network. A dynamic framework is thus needed to accurately allocate responsibility for negative social, environmental, or economic impacts in a network. 

\medskip 

This need to hold players responsible for the negative impacts that occur in their supply chain is best illustrated by the example of greenhouse gas emissions. A significant proportion of the carbon emissions associated with a given firm takes place elsewhere in its supply chain - for example, during the manufacturing process that has been outsourced to another country, during the extraction phase of raw materials, or during the end-of-life disposal of a product \parencite{hertwich2018growing}. While there seems to be a consensus that players should bear some responsibility for these Scope 3 emissions, which include all indirect emissions from a player's value chain \parencite{GHGprotocol}, it is not clear to \emph{what extent} players should be held responsible or even liable.

\medskip 

In this paper, we provide such a framework to quantify the responsibility of players in a dynamic supply chain or network for the impacts of the actions of players in their network. We describe how responsibility is embedded in the network and how it diffuses from one player to another, based on how players are connected to each other at each moment in time and on the direct impacts of each player. We aim to develop a model that is general enough to be applicable to responsibility for \emph{any} kind of impact that needs to be reallocated. We are not just considering greenhouse gas emissions, but can also think of for example systemic risk, biodiversity loss, soil degradation or (virtual) water use. 

\medskip 

The responsibility we assign to a given player depends on (i) the underlying network structure of all players at each time \(t\), (ii) some initial state of responsibility, and (iii) the rate at which each player generates impact (e.g., GHG emissions) over time, represented by a continuous function called the impact map \(s: \mathbb{R}_+ \to \mathbb{R}^n\). Analysing impacts per unit of time rather than relying on averaged or aggregated values allows for a more precise allocation. This approach captures seasonal variations and can account for players involved in multiple activity streams of varying intensity at different times. Consequently, our responsibility measure is a continuous function that assigns a certain amount of responsibility to each player \emph{at any time \(t\).}
Responsibility for some impacts starts initially at the source, i.e., at the player who generated the impact, and then flows to the other players according to who benefits directly and indirectly from these impacts. This flow is governed by the Laplacian matrix, the discrete analogue of the Laplace operator, a second-order differential operator best known from the heat equation. According to the second law of thermodynamics, the amount of heat that flows from a source to another point depends on the temperature difference and the conductivity of the material between them. In this paper, responsibility flows in a similar manner: it is proportional to the difference in direct impact between two nodes, and the weight of the walks connecting them. Although concepts of physics in general and the heat equation in particular have been applied to a wide range of problems, for example in theoretical biology \parencite{mirzaev2013laplacian} or in general economic problems \parencite{dedomenico_2017} \parencite{shuman2013emerging} \parencite{masuda.etal_2017} \parencite{battiston.etal_2012}, we take a novel approach by adopting it to allocate responsibility for impacts that occur throughout a network or supply chain. 

\medskip 

We derive the differential equation for our responsibility measure. In particular, responsibility for all impacts generated must be subject to diffusion, and the amount that diffuses to each player depends only on how much they benefit from those impacts. Furthermore, we show that the responsibility measure satisfies the properties of (i) efficiency, which avoids double counting; (ii) linearity, which prevents strategic manipulation of responsibility by merging or splitting; (iii) the independent player property, ensuring that players who do not benefit from anyone else's emissions do not bear responsibility for them; and, finally, (iv) symmetry, which ensures that two players with the same role in the network receive the same indirect responsibility. In addition, the responsibility measure also exhibits temporal consistency: evolving the system from an initial time to some later time can be broken down into intermediate steps without loss of generality. 

\medskip 

In the context of greenhouse gas emissions, various methods to attribute responsibility have been put forward in the literature. For a review of allocations of carbon emissions, we refer to \textcite{zhou2016carbon}. An approach that is often applied is by using input-output models, which has led to the notions of consumer-based accounting \parencite{eder1999environmental,munksgaard2001co2}, income-based accounting \parencite{marques2012income} or production based accounting \parencite{franzen2018consumption}. Another related work is by \textcite{vandenende2023network} who assign responsibility for greenhouse gas emissions in a supply chain based on the discounted sum over walks between all players. The present paper builds on the same assumption that responsibility depends both on a player's direct emissions and the role this player has in the network. However, we extend the analysis by allowing for dynamic networks and by considering a continuous impact function instead of aggregating emissions over a longer period. In addition, we take a more general approach that goes beyond greenhouse gas emissions, allowing to allocate responsibility for any kind of impact. Additional related work in the context of environmental responsibility includes that of \textcite{chassagneux2023modelling}, who model carbon markets such as the EU-ETS by using stochastic differential equations, or by \textcite{gopalakrishnan2021incentives} \parencite{Gopalakrishnan2021ConsistentChains}
who adopt a cooperative game theoretical approach to allocate responsibility. Cooperative game theoretical methods have been applied more generally to design fair allocation mechanisms for environmental costs and benefits, particularly in transboundary pollution control and resource management \parencite{chander1995}, river management \parencite{ni2007sharing,gomez2013sharing, alcalde2015sharing,brink2018polluted} or in coal-fired power plants \parencite{zheng2015}.  Solutions such as the Shapley value \parencite{shapley1953value} and the nucleolus \parencite{schmeidler1969nucleolus} have been used to ensure that all parties share the environmental burdens and benefits fairly, based on their contributions and impacts. 

\medskip 

This paper distinguishes itself from the existing literature on responsibility allocation by exploiting the dynamic nature of network structures. Dynamic or temporal networks are studied by for example \textcite{snijders2001statistical}, who introduces stochastic actor-oriented models in which the evolution of networks is modelled as a consequence of individual decisions of players. This work is extended by \textcite{niezink2019no} and applied by \textcite{greenan2015diffusion} to study the diffusion of innovations in dynamic networks. Additionally, \textcite{li2017fundamental} demonstrates that temporal networks exhibit fundamental advantages compared to static networks in terms of controllability and reachability. For a formal mathematical treatment of temporal networks, we refer to \textcite{masuda2016guide}.

\medskip 

We maintain broad applicability as we aim to adopt a general approach that can be applied to a wider range of impacts for which responsibility needs to be allocated in order to achieve the corresponding (reduction) goals accordingly. Our paper also acts as a bridge between, for example, the seminal work on the carbon footprint of nations by \textcite{hertwich2009carbon} and concepts such as the biodiversity footprint, which assesses the effect of operations on the biodiversity \parencite{yue2023analysis,bjelle2021trends}, or the material footprint of nations which focuses not just on carbon emissions but on all materials embedded in products \parencite{wiedmann2015material}. Closely related to this is the concept of true pricing, which aims to include not only the embedded carbon emissions in the consumer price of a product, but also factors such as biodiversity loss or an unfair wages for workers \parencite{hendriks2023true}. Another example can be found in the context of virtual water trade, which can be seen as an indicator that measures the amount of water that is embedded in goods or services as a result of water-intensive (production-)processes that take place along the supply chain \parencite{sartori2017modeling}. Though the need for such a measure is clear, there is still much debate about its theoretical foundations \parencite{reimer2012economics}, which is a gap we try to fill. 

\medskip 

The paper is structured as follows. In the next section, we explain all the ingredients and mathematical tools that we will use throughout the paper. In Section~\ref{sec:allocating}, we construct the responsibility measure and describe how it diffuses through the network. We derive a differential equation and initial condition that characterizes this evolution. Subsequently, in Section \ref{sec:prop}, we establish a set of desirable properties of the dynamic responsibility measure and relate it to the Shapley value. To facilitate the practical use of our model, we show in Section~\ref{sec:approx} approximating the measure does not compromise its properties. Finally, in Section~\ref{sec:discussion}, we conclude with a discussion of possible extensions and applications including to multi-layered networks or hypergraphs.

\section{Preliminaries}
We allocate responsibility for impacts that occur throughout the network. In this section, we outline the key theories and concepts that are needed to establish this allocation method. 

\subsection{Graphs and adjacency matrices}
A \emph{directed network} \(G\) is an ordered pair \( G = (N, E)\) with \(N = \{1,\ldots,n\}\) a set of vertices and \(E \subseteq N \times N\) a set of ordered pairs of vertices, called the \emph{edges}. For a \emph{weighted} network, a nonnegative weight \(A_{ij}\) is associated to each edge \((i, j) \in E\). Whenever the network is \emph{dynamic}, the adjacency matrix changes over time. We thus denote by \(A^{(t)}\) the corresponding adjacency matrix at time \(t\). In our setting, we call \(N\) the set of \emph{players}. Each edge \((i, j)\), with \(i\) the source node and \(j\) the terminal node, can be interpreted as an interaction between players \(i\) and \(j\), with \(i\) benefiting from the actions of player \(j\), and consequentially taking partial responsibility for the actions of this player. For example, \(A^{(t)}_{ij}\) could represent how much of the sales of \(j\) at time \(t\) go to \(i\).

\medskip 

From this normalization, it follows that the adjacency matrix \(A^{(t)}\) is \emph{column stochastic}: we have \(\sum_{i \in N} A^{(t)}_{ij} = 1\) for every \(j \in N\) at all \(t \in \mathbb{R}_{+}\) and each entry is non-negative. An example of a simple static network and its corresponding adjacency matrix can be seen in Figure \ref{fig:3players}, where player \(a\) uses \(0.1\) units of her production for herself, sells \(0.3\) units to player \(b\) and \(0.6\) units to player \(c\). Player \(b\) only sells to player \(c\), and player \(c\) sells most of her production, \(0.8\) units, to player \(b\) while she keeps \(0.2\) units for herself. We let \(D^{(t)} \in \bbR^{N\times N}\) be the diagonal \emph{degree matrix} associated to \(A^{(t)}\), where the \(i\)-th entry is given by the sum of weights of its incoming links, or equivalently, \(D^{(t)}_{ii} = \sum_{j\in N} A^{(t)}_{ji}\). Remark that for column stochastic matrices, this degree matrix is equal to the identity matrix.

\begin{figure}[h!!]
\centering 
\hspace{0.08\textwidth}
\begin{subfigure}[h]{0.3\textwidth}
\begin{tikzpicture}
    \node[shape=circle,draw=black] (a) at (0,2) {$a$};
    \node[shape=circle,draw=black] (b) at (-1.8, 0) {$b$};
    \node[shape=circle,draw=black] (c) at (1.8,0) {$c$};
    \path [->] (b) edge[above left]  node {$0.3$} (a);
    \path [->] (a) edge[loop above]  node {$0.1$} (a);
    \path [->] (c) edge[above right] node {$0.6$} (a);
    \path [->] (b) edge[bend left = 20, above] node {$0.8$} (c);
    \path [->] (c) edge[bend left = 20, below] node {$1$} (b);
    \path [->] (c) edge[loop below] node {$0.2$} (c);
\end{tikzpicture}
\end{subfigure}
\begin{subfigure}[h]{0.49\textwidth}
\[
A = \begin{pmatrix} 
0.1 & 0 & 0  \\
0.3 & 0 & 0.8  \\
0.6 & 1 & 0.2  \\
\end{pmatrix}
\]
\end{subfigure}
\caption{A network and its corresponding adjacency matrix.}
\label{fig:3players}
\end{figure}

\subsection{Laplacian matrices}
The \emph{Laplacian matrix} \(\mathcal{L} \in \bbR^{N\times N}\) of a network \(G\) is defined as \(\mathcal{L} = D - A\) with \(A\) the adjacency matrix and \(D\) the diagonal (in)degree matrix of \(A\). The Laplacian matrix can be seen as a natural discrete extension on graphs of the Laplace operator, the divergence of the gradient of a function, which plays an important role in explaining various phenomena in physics, such as the diffusion equation of the heat flow or electric potentials. Moreover, the Laplacian matrix relates to many spectral properties of the graph, such as the number of spanning trees \parencite{hammer1996laplacian}. Since we always consider the Laplacian of column stochastic matrices in this model, the Laplacian at time \(t \in \mathbb{R}_{+}\) is defined as \(\calL^{(t)} = I - A^{(t)}\). 

\subsection{Impact map and initial condition}

Having defined the network, we now want to describe what generates responsibility. We model it by an \emph{impact map} \(s: \mathbb{R}_+ \to \mathbb{R}^N\), a continuous function that gives, for any time \(t \in \mathbb{R}_+\), the amount of impact \(s(t) \in \mathbb{R}^N \) created by each player for an infinitesimal duration. For example, the impact map could be interpreted as the direct emission of greenhouse gases per unit of time. Describing the impact in this manner allows to take into account that players have some fluctuating impact and, for instance, pollute more in some seasons, or hours, than others. Their activities could consist of various kinds of operations that are not necessarily equally polluting. Then, players that are connected to a given player for some given time period, are only held responsible for the impacts that occurred during this exact time period. For simplicity and intuition, one can consider \(s\) to be a continuous function if the time scale we are considering is finite. However, keep in mind that we do allow for e.g. policy shocks that result in a discontinuous impact function, as long as \(s\) remains integrable. Finally, we specify an initial condition stating that the responsibility at time \(t=0\) is given by some \(f \in \mathbb{R}^N\). This allows us, for example, to embed historical emissions. 

\section{Allocating responsibility}
\label{sec:allocating}

We denote by \(\mathcal{A}^N\) the set of column stochastic matrices of size \(\lvert N \rvert \times \lvert N \rvert\) and by \(\mathcal{C}(\mathbb{R}_+, \mathbb{R}^N)\) the set of continuous maps from \(\mathbb{R}_+\) to \(\mathbb{R}^N\), with \(\mathbb{R}_+\) denoting the set of nonnegative real numbers. We aim to define a \emph{responsibility allocation} 
\[
\rho\colon \mathcal{C} \left( \mathbb{R}_+, \mathcal{A}^N \right) \times \bbR^N \times \mathcal{C} \left( \bbR_+, \bbR^N \right) \longmapsto \mathcal{C} \left( \bbR_+, \bbR^N \right),
\]
that maps each triplet consisting of a dynamic network, an (initial) state and a continuous source onto a continuous function that assigns to each player a responsibility at time \(t\). When there is no risk of confusion, we use the slight abuse of notation to write \(\rho = \rho(A, f, s)\) to denote the images of \(\rho\) as well as \(\rho\) itself. Hence, \(\rho[t]\) is short for \(\rho(A, f, s)[t]\). Moreover, throughout this section, we set \(t\) to be a fixed but arbitrary positive real number. 

\medskip 

To find the appropriate \(\rho[t]\), we evaluate the effect of one unit of impact taking place somewhere in the network, and see how the responsibility for this unit of impact spreads to the other players. In other words, we are interested in the \emph{evolution} of responsibility. 
Particularly, we evaluate the partial differential of the responsibility map, \(\partial_t \rho\). Suppose that player \(j \in N\) is connected in a network \(G\), and has an initial responsibility of one unit of impact, while the other players carry zero responsibility.  The responsibility of player \(i\in N\) evolves according to three effects: 

\begin{equation}
\label{eq:evolution} 
\partial_t \rho_i[t] = - \delta_{ij} + A^{(t)}_{ij} + s_i(t).
\end{equation}

Each term in this expression reflects a different mechanism. First, there is the outflow: the responsibility of player \(j\) begins to flow outward, modeled by the term \(\delta_{ij}\). This reflects the fact that only player \(j\) starts with responsibility, so only this player can lose responsibility initially. Second, there is the inflow. The responsibility outflowing player \(j\) is distributed to other players based on the netwerk structure. Recall that our adjacency matrices are column stochastic, that is, \(\sum_{i \in N} A^{(t)}_{ij} = 1\) and that \(A^{(t)}_{ij}\) captures how much player \(i\) benefits from player j, so responsibility flows accordingly. If a player benefits more from the impacts of a player, it is assigned more responsibility for these impacts. Finally, there is the generation of responsibility: over time, players also generate new responsibility by creating new impacts, represented by the source term \(s_i(t)\).

\medskip

Thus, the responsibility for the impact created by player \(j\) flows to all players according to how much they benefit from \(j\)'s actions. From the perspective of player \(j\), the entire impact is subject to diffusion and begins to flow outwards. If \(j\) benefits from its own actions, part of the responsibility flows back. The sign of \([\partial_t \rho]_i \) indicates whether the responsibility of player \(i\) is increasing or decreasing. 
For player \(i \neq j\), where \(\delta_{ij} = 0\), the evolution of responsibility is never negative, i.e., the responsibility cannot decrease. On the other hand, assuming \(s(t) = 0\), if we let \(i = j\) and \(A^{(t)}_{jj} = 0\) (player \(j\) does not benefit from itself), responsibility decreases steeply with \(\partial_t \rho_j[t] = -1 \). On the contrary, for \(A^{(t)}_{jj} = 1\), we obtain \(\partial_t \rho_j[t] = 0\), indicating constant responsibility. This occurs when no other player benefits from \(j\), leaving \(j\) fully responsible for the unit of impact it created.

\begin{example}
\label{ex:dtrho}
Consider the graph of Figure~\ref{fig:3players} at time \(t\). We let player \(a\) be responsible for \(1\) unit of impact, while players \(b\) and \(c\) carry no initial responsibility. We have \(A_{aa}^{(t)} = 0.1\), \(A_{ba}^{(t)} = 0.3\) and \(A_{ca}^{(t)} = 0.6\): player \(a\) benefits for 0.1 of its own impacts, while player \(b\) and player \(c\) benefit for respectively 0.3 and 0.6. None of the players create any new impacts at time \(t\), that is, \(s(t) = 0\) for all players. We then find
\[
\partial_t \rho_a[t] = 0.1 - 1 = -0.9, \qquad \qquad \partial_t \rho_b[t] = 0.3, \qquad \qquad \partial_t \rho_c[t] = 0.6
\]
We can indeed see that the responsibility of player \(a\) decreases steeply. How much flows to every player, including player \(a\) itself, depends on how much each player benefits from these impacts. 
\end{example}

Note that the flow of responsibility is in a direction opposite of the arrows of Figure~\ref{fig:3players}. Moreover, remark that \(\partial_t \rho_i\) is the \emph{evolution} of responsibility of player \(i\), not the responsibility itself. If we find \(\partial_t \rho_i = -1\) for some player \(i\), this implies that at that infinitesimal time step, her responsibility is decreasing steeply. However, for any \(t' > t\), the current responsibility of the player has changed, as well as the derivative, hence it does not imply that all of her responsibility is transferred to the other players. Over time, the state evolves, and the rate of change of responsibility adjusts accordingly. 

\medskip
Equation~\eqref{eq:evolution} emphasizes the graph-theoretic nature of the diffusion. Regardless of how much responsibility a player has, the diffusion is only determined by who the players are connected to and how strong those connections are. From just this equation, and the fact that the partial derivative with respect to time \(\partial_t\) is linear, we can determine how the full responsibility measure evolves over time. 

\begin{theorem}\label{thm:evol}
Let \(\calL\) be the Laplacian matrix of the network where the players' current responsibility is given by \(f \in \mathbb{R}^N\). The evolution of the responsibility allocation at time \(t\) satisfying Equation~\eqref{eq:evolution} is uniquely determined by 
\begin{equation}\label{eq: theorem}
\begin{cases}
    \partial_t \rho[t] = - \calL^{(t)} \rho[t] + s(t) \\
    \rho[0] = f.
\end{cases}
\end{equation}
\end{theorem}

\begin{proof}
Let \(\mathbf{1}^{i}\) denote the vector of size \(N\) whose element \(j\) is  \(1\) and that is zero everywhere else. By linearity of \(\partial_t\) and from the source term in Equation~\eqref{eq:evolution}, 
we can write for any current \(\rho[t] = \sum_{j \in N} \rho[t]_j \mathbf{1}^{j}\) 
\begin{equation}\label{eq: linear-decomp}
\partial_t \rho[t] = \sum_{j \in N} \rho[t]_j \partial_t \mathbf{1}^j + s(t). 
\end{equation}
Let us temporarily focus on the general term of the sum \(\partial_t \mathbf{1}^j\). Using Equation~\ref{eq:evolution}, we get that the \(i\)th entry equals \( \left[ \partial_t \mathbf{1}^j \right]_i = A^{(t)}_{ij} - \delta_{ij} \), then the vector is given by 
\[
\partial_t \mathbf{1}^j = A^{(t)} \mathbf{1}^j - I \mathbf{1}^j = - \mathcal{L}^{(t)} \mathbf{1}^j. 
\]
Plugging this back into Equation~\eqref{eq: linear-decomp} yields 
\[
\partial_t \rho[t] = \sum_{j \in N} \rho[t]_j \left( - \mathcal{L}^{(t)} \mathbf{1}^j \right) + s(t) = - \mathcal{L}^{(t)} \sum_{j \in N} \rho[t]_j \mathbf{1}^j + s(t) = - \mathcal{L}^{(t)} \rho[t] + s(t). 
\]
We invoke the Picard-Lindel{\"o}f Theorem to conclude on the existence and unicity of \(\rho[t]\) satisfying any initial condition \(\rho[0] = f\) and Equation~\eqref{eq: theorem}. 
\end{proof}

The solution is given by
\[
\rho[t] = \Phi(t, 0) f + 
\int_0^t \Phi(t, \tau) s(\tau) \mathrm{d}\tau
\]
with \(\Phi(t, t_0)\) being the matrix given by the Peano-Baker series \parencite{baake2011peano}:
\[ \begin{aligned} 
\Phi(t, t_0) = \sum_{k \geq 0} F_k(t), \qquad \text{with} \qquad F_{k+1}(t) = \int_{t_0}^t \calL(\tau) F_k(\tau) \mathrm{d} \tau, \quad \text{and} \quad F_0(t) = I. 
\end{aligned} \]
Since \(\mathcal{L}(t_1) \mathcal{L}(t_2) \neq \mathcal{L}(t_2) \mathcal{L}(t_1)\), we have to ensure that matrix multiplications occur in the correct order. The state transition matrix \(\Phi(t, t_0)\) does exactly this: it provides a way to handle the non-commutativity of Laplacian matrices at different time steps. This situation is analogous to the time-ordering operator \(\mathcal{T}\) in quantum field theory \parencite{weinberg1995quantum}. 

\section{Properties}\label{sec:prop}
 To gain more intuition for this responsibility measure and to enhance its applicability in contexts such as legal or policy assessment, we further analyse the dynamic responsibility measure and show that it satisfies a set of desirable properties. We begin by formalising the notion of individual responsibility over time and introducing a decomposition that reveals its internal structure of direct and indirect responsibility. 

\medskip 

We denote by \(\rho_i[t]\) the responsibility of player \(i \in N\) at time \(t \in \mathbb{R}_+\). In particular, it is the solution to 
\begin{equation}
\label{eq: individual-responsibility}
\partial_t \rho_i[t] = - \calL^{(t)}_{ii} \rho_i[t] - \sum_{j \neq i} \calL^{(t)}_{ij} \rho_j[t] + s_i(t), \qquad \rho_i[0] = f_i. 
\end{equation}
It can be decomposed in two parts,
\begin{equation*}
\rho_i[t] = \mu_i(t) + \nu_i(t), 
\end{equation*}
where \(\mu_i(t)\) is the direct responsibility coming from its impact \(s_i\) and initial responsibility \(f_i\), and \(\nu_i(t)\) is the responsibility coming from its interaction with other nodes, starting at \(\nu_i(0) = 0 \). For a given player \(i \in N\), its direct responsibility at time \(t \in \mathbb{R}_+\) is given by the solution to the following Cauchy problem, 
\begin{equation}\label{eq: direct-responsibility-definition}
\partial_t \mu_i(t) = - \calL^{(t)}_{ii} \rho_i[t] + s_i(t), \qquad \mu_i(0) = f_i,
\end{equation}
obtained from Equation~\eqref{eq: individual-responsibility} by only considering its own responsibility and impact. The contribution coming from the activity of all the other players, which we call the \emph{indirect responsibility} of this player, is defined as the solution to
\begin{equation}\label{eq: indirect-responsibility-definition}
\partial_t \nu_i (t) = - \sum_{j \neq i} \calL^{(t)}_{ij} \rho_j(t), \qquad \nu_i(0) = 0.
\end{equation}
Hence, the indirect responsibility is indeed generated by the diffusion of the responsibility of ones (indirect) connections, weighted by their proximity, i.e., the strength of their relation. 

\medskip 

First, we prove that \(\rho[t]\) takes only nonnegative values. 

\begin{lemma}\label{lemma: non-negativity}
    Let \(f \in \mathbb{R}^n_+\) be the initial condition of the Cauchy problem~\eqref{eq: theorem}, and let \(s \in \mathcal{C}(\mathbb{R}_+, \mathbb{R}_+^n)\) take only nonnegative values. Then for all \(t \in \mathbb{R}_+\), we have \(\rho[t]\in \mathbb{R}_+^n\). 
\end{lemma}

\begin{proof}
    We prove this statement by contradiction. Assume that there exists \(t > 0\) such that there exists a player \(k \in N\) with \(\rho_i[t] < 0\). Therefore, we have that the set \(T\) of real numbers \(t \in \mathbb{R}_+\) such that there exists a player \(k \in N\) for which 
    \[
    \rho_k[t] = 0 \qquad \text{ and } \qquad \partial_t \rho_k[t] < 0 
    \]
    is nonempty and bounded by below. Hence, \(t_\star = \inf T\) is well-defined. Denote by \(i\) a player for which \(\rho_i[t_\star] = 0\) and \(\partial_t \rho_i[t_\star] < 0\). Developing the time derivative using Equation~\eqref{eq: individual-responsibility} yields
    \[
    \partial_t \rho_i [t_\star] = - \calL_{i i}(t_\star) \rho_i[t_\star] - \sum_{j \neq i} \calL_{i j}(t_\star) \rho_j[t_\star] + s_i(t_\star). 
    \]
    Because \(\rho_i[t_\star] = 0\), we can simplify as 
    \[
    \partial_t \rho_i[t_\star] = - \sum_{j \neq i} \calL_{i j} (t_\star) \rho_j[t_\star] + s_i(t_\star) = \sum_{j \neq i} A_{i j}(t_\star) \rho_j[t_\star] + s_i(t_\star). 
    \]
    Because of the definition of \(i\), we have that \(\rho_j[t_\star]\) is nonnegative for all \(j \neq i\). Together with the column stochasticity of \(A\) and non-negativity of \(s\) at any time, we observe that the right-hand side is nonnegative, while the left-hand side is negative, which is absurd. 
\end{proof}

Next, we discuss some properties of our responsibility measure. The first property regards efficiency, i.e., the total responsibility equals the sum of the impacts. 

\begin{proposition}[Efficiency]
    Let \(\calL\) be a time-varying Laplacian matrix, let \(f \in \mathbb{R}_+^*\) be the initial responsibility, and let \(s \in \mathcal{C}(\mathbb{R}_+, \mathbb{R}_+^n)\) be the impact map. Then, for all \(t \in \mathbb{R}_+\), we have
    \[
    \lVert \rho[t] \rVert_1 = \lVert f \rVert_1 + \int_0^t \lVert s(\tau) \rVert_1 \mathrm{d}\tau. 
    \]
    \label{prop:efficiency}
\end{proposition}

\begin{proof}
    Let \(t \in \mathbb{R}_+\) be arbitrary. First, by Lemma~\ref{lemma: non-negativity}, we know that \(\rho[t]\) is non-negative. Hence, its \(\ell^1\)-norm satisfies \(\lVert \rho[t] \rVert_1 = \mathbf{1}^\top \rho[t]\). Let us investigate the time derivative of this norm. We have
    \[
    \partial_t \lVert \rho[t] \rVert_1 = \partial_t \mathbf{1}^\top \rho[t] = \mathbf{1}^\top \partial_t \rho[t]. 
    \]
    We plug Equation~\eqref{eq: theorem} into this to obtain
    \[
    \partial_t \lVert \rho[t] \rVert_1 = \mathbf{1}^\top \left( - \calL(t) \rho[t] + s(t) \right) = - \mathbf{1}^\top \calL(t) \rho[t] + \mathbf{1}^\top s(t). 
    \]
    Because the adjacency matrices are column stochastic at any time, we have that \(\mathbf{1}^\top \calL(t)\) is the null vector. The non-negativity of \(s\) gives that \(\mathbf{1}^\top s(t) = \lVert s(t) \rVert_1\), then
    \[
    \partial_t \lVert \rho[t] \rVert_1 = \lVert s(t) \rVert_1. 
    \] 
    Integrating on both sides yields
    \[
    \lVert \rho[t] \rVert_1 = \lVert \rho[0] \rVert_1 + \int_0^t \lVert s(\tau) \rVert_1 \mathrm{d}\tau = \lVert f \rVert_1 + \int_0^t \lVert s(\tau) \rVert_1 \mathrm{d} \tau,
    \]
    which concludes the proof. 
\end{proof}

Efficiency ensures that every unit of impact is assigned exactly once and to only one player: the total responsibility equals the total impact. This avoids \emph{double counting}, a concept well-known in the attribution of responsibility for carbon emissions (see, for example \textcite{schneider2019double}) Double counting is particularly problematic when the reduction of a negative impact is attributed to more than one player, even though the actual (physical) reduction occurs only once. This leads to inaccurate and inflated reported reductions, undermining the accuracy of climate action efforts. This principle of double-counting is also applicable to other kinds of impacts for which we want to allocate responsibility. Therefore, in order to have an appropriate accounting mechanism, we want the property of efficiency to be satisfied. 

\medskip 

The next proposition is linearity, which ensures that if a responsibility is allocated within the same network but for two different impact maps and initial conditions, the total responsibility is equal to the sum of the responsibilities calculated for each impact map and initial condition individually.  

\begin{proposition}[Linearity]
    The responsibility value is linear with respect to the current responsibility and the impact function. 
\end{proposition}

\begin{proof}
Let \(t \in \mathbb{R}_+\), \(\lambda \in \mathbb{R}\) let \(\hat{f}, \tilde{f} \in \mathbb{R}_+^n\) be two vectors of prior responsibility and let \(\hat{s}, \tilde{s} : N \to \mathbb{R}_+\) be two impact maps. We want to show that 
\[
\rho \left( \lambda \hat{f} + \tilde{f}, \lambda \hat{s} + \tilde{s} \right) [t] = \lambda \rho \left( \hat{f}, \hat{s} \right) [t] + \rho \left( \tilde{f}, \tilde{s} \right) [t]. 
\]
Let rewrite \(\hat{\rho} = \rho \left( \hat{f}, \hat{s} \right)\) and \(\tilde{\rho} = \rho \left( \tilde{f}, \tilde{s} \right)\). By linearity of the differential operator \(\partial_t\), we have that 
\[
\partial_t \left( \lambda \hat{\rho} [t] + \tilde{\rho} [t] \right) = \lambda \partial_t \hat{\rho} [t] + \partial_t \tilde{\rho} [t]. 
\]
Next, we replace each \(\partial_t \rho\) by their formula, to get
\[ \begin{aligned} 
\partial_t \left( \lambda \hat{\rho} [t] + \tilde{\rho} [t] \right) & = \lambda \left( - \mathcal{L}^{(t)} \hat{\rho}[t] + \hat{s}(t) \right) - \mathcal{L}^{(t)} \tilde{\rho}[t] + \tilde{s}(t) \\
& = - \mathcal{L}^{(t)} \left( \lambda \hat{\rho} + \tilde{\rho} \right) [t] + \left( \lambda \hat{s} + \tilde{s} \right)(t).
\end{aligned} \]
This defines the following Cauchy problem, 
\begin{equation}\label{eq: linear}
\begin{cases}
    \partial_t \overline{\rho}[t] = - \calL^{(t)} \overline{\rho}[t] + \left( \lambda \hat{s} + \tilde{s} \right)(t), \\
    \overline{\rho}[0] = \lambda \hat{f} + \tilde{f},
\end{cases}
\end{equation}
of which \(\lambda \hat{\rho} + \tilde{\rho}\) is a solution as shown by the computations above. By definition, \(\rho \left( \lambda \hat{f} + \tilde{f}, \lambda \hat{s} + \tilde{s} \right)\) is a solution to this Cauchy problem as well. By the Picard-Lindel{\"o}f Theorem, the solution to Equation~\eqref{eq: linear} is unique, hence both solutions coincide.
\end{proof}

 Linearity is desirable because it makes the responsibility allocation invariant under relabeling, grouping, or artificial splitting of players. For example, if two firms with distinct emissions and network positions choose to jointly report their activity or, conversely, a single firm reports its emissions as coming from its subsidiaries, the total responsibility attributed to them remains unchanged, as long as the actual emissions and network structure stay the same. Linearity thus prevents “smart accounting” strategies where players attempt to reduce their responsibility through purely formal restructuring, without altering their emissions or the network structure. Linearity also allows us to isolate and study the effect of any individual player. This allows, for example, a country or firm to assess its cumulative responsibility over a given period, or to evaluate how a change in its emissions path  propagates through the system. Moreover, linearity enables the model to assign responsibility for multiple pollutants: for example, responsibilities for the emissions of carbon dioxide and of methane can be computed separately and then aggregated according to some policy-relevant weights. 

In addition to linearity, our allocation also exhibits temporal consistency, a consequence of the semigroup property of the time-evolution operators.
Specifically, evolving the system from an initial time to a later time can be broken down into intermediate steps without loss of generality. That is, if we evolve from \(t_0\) to \(t_1\) and then from \(t_1\) to \(t_2\), we obtain the same results as when we would evolve directly from \(t_0\) to \(t_2\). A consequence of this is that the responsibility at any given time serves as the initial condition for the next time step. Thus, the impact function at time \(t\) is embedded in the initial condition of later times. In other words, once we know the responsibility at some time \(t\), we can compute the responsibility at time \(t^{\prime} > t\) without requiring knowledge about the full history of the system.

\medskip 

We next evaluate the responsibility of players who do not benefit from the actions of any other player at any time \(t\). We call such players \emph{independent players}.

\begin{definition}[Independent player property]
    Let \(t\) be a positive real number. We say that the player \(i \in N\) is an \emph{independent player} during \([0, t^*]\) if we have \(A_{ij}(t) = 0\) for all \(j \in N\) and \(t \in [0, t^*]\). 
\end{definition}

In a network, independent players do not have any outgoing edges. In Figure~\ref{fig:indepsymm} we see that player \(d\) is an independent player, as she does not benefit from any player. 

\begin{figure}[h!!]
\centering 
\hspace{0.08\textwidth}
\begin{subfigure}[h]{0.4\textwidth}
\begin{tikzpicture}
\begin{scope}[rotate = -90]
    \node[shape=circle,draw=black] (a) at (0,0) {$a$};
    \node[shape=circle,draw=black] (b) at (4, 0) {$b$};
    \node[shape=circle,draw=black] (d) at (2, 4.5) {$d$};
    \node[shape=circle,draw=black] (c) at (2,3) {$c$};
    \path [->] (a) edge[loop left]  node {\(\nicefrac{1}{4}\)} (a);
    \path [->] (a) edge[bend left =20, below]  node[sloped,left] {\(\nicefrac{1}{3}\)} (b);
    \path [->] (a) edge[bend right = 20, above] node[] {\(\nicefrac{1}{2}\)} (c);
    \path [->] (b) edge[bend left = 20, below]  node[left] {\(\nicefrac{1}{3}\)} (a);
    \path [->] (b) edge[loop right]  node {\(\nicefrac{1}{4}\)} (b);
    \path [->] (b) edge[bend left = 20, above] node[right] {\(\:\: \nicefrac{1}{2}\)} (c);
    \path [->] (c) edge[bend right = 20, above] node[] {\(\nicefrac{5}{12}\)} (a);
    \path [->] (c) edge[bend left = 20, above] node [right]{\(\:\: \sfrac{5}{12}\)} (b);
    \path [->] (c) edge[left]  node[below] {\(1\)} (d);
\end{scope}
\end{tikzpicture}
\end{subfigure}
\begin{subfigure}[h]{0.49\textwidth}
\[
A = \begin{pmatrix} 
\nicefrac{1}{4} & \nicefrac{1}{3} & \nicefrac{1}{2} & 0  \\
\nicefrac{1}{3} & \nicefrac{1}{4} & \nicefrac{1}{2} & 0  \\
\nicefrac{5}{12} & \nicefrac{5}{12} & 0 & 1  \\
0 & 0 & 0 & 0\\

\end{pmatrix}
\]
\end{subfigure}
\caption{
A network of players \(\{a, b, c, d\}\), along with its corresponding adjacency matrix. Players \(a\) and \(b\) are symmetric players with respect to each other, and player \(d\) is an independent player.}
\label{fig:indepsymm}
\end{figure}

This leads to the following result about independent players.

\begin{proposition}\label{prop: independent-player}
    The indirect responsibility of an independent player is zero. 
\end{proposition}

\begin{proof}
    Let \(t^*\) be a positive real number, let \(i \in N\) be an independent player during \([0, t^*]\) and \(t \in [0, t^*]\) be arbitrary. From equation~\eqref{eq: indirect-responsibility-definition}, we have that the indirect responsibility of \(i\) at time \(t\) is the solution of 
    \[
    \partial_t \nu_i(t) = - \sum_{j \neq i} \mathcal{L}^{(t)}_{ij} \rho_j [t], \qquad \nu_i(0) = 0. 
    \]
    We unfold the Laplacian matrix to get 
    \[
    \partial_t \nu_i (t) = \sum_{j \neq i } A^{(t)}_{ij} \rho_j[t]. 
    \]
    Because \(i\) is an independent player, we have that \(A^{(t)}_{ij} = 0\) for all \(j \in N\) and \(t \in [0, t^*]\), therefore 
    \[
    \partial_t \nu_i(t) = 0.
    \]
    Because the initial condition are \(\nu_i(t) = 0\), and its time derivative is zero, the indirect responsibility of an independent player is always zero. 
\end{proof}

\begin{corollary}\label{cor: independent-player}
    The responsibility of an independent player is given by
    \[
    \rho_i[t] = e^{-t} f_i + \int_0^t e^{\tau - t} s_i(\tau) \mathrm{d} \tau. 
    \]
\end{corollary}

\begin{proof}
    Let \(t^*\) be a positive real number, let \(i \in N\) be an independent player during \([0, t^*]\) and \(t \in [0, t^*]\) be arbitrary. Using Proposition~\ref{prop: independent-player}, we know that the responsibility of \(i\) is equal to its direct responsibility, and from Equation~\eqref{eq: direct-responsibility-definition} we have  
    \[
    \partial_t \rho_i[t] = - \mathcal{L}^{(t)}_{ii} \rho_i [t] + s_i(t). 
    \]
    Because \(i\) is independent and \(A^{(t)}\) is column stochastic, we have that 
    \[
    \partial_t \rho_i [t] = s_i(t) - \rho_i [t]. 
    \]
    To check that the formula claimed in the statement above is valid, we compute its time derivative: 
    \[
    \partial_t \rho_i[t] = - e^{-t} f_i + e^{-t} \left( - \int_0^t e^\tau s_i(\tau) \mathrm{d}\tau + e^t s_i(t) \right) = s_i(t) - e^{-t} f_i - \int_0^t e^{\tau - t} s_i(\tau) \mathrm{d} \tau, 
    \]
    which is indeed equal to \(s_i(t) - \rho_i[t]\) as desired. 
\end{proof}

In other words, since \(e^{-t} \leq e^{\tau - t} \leq 1\) for positive values of \(t\), the responsibility of independent players is lower than the sum of their impacts. This is explained by the fact that these players do not benefit from anyone else, but others do benefit from them, so part of the responsibility has flowed to these other players. Next, we discuss the case when two players are identical regarding their incoming and outgoing links. We call these players \emph{symmetric players}.

\medskip 

Let \(\sigma_{ij}\) be the bijection \(\sigma_{ij}: N \to N\) that acts like the identity on \(N \setminus \{i, j\}\) and swaps \(i\) and \(j\). For any matrix \(X\), we denote by \(\sigma_{ij} \left( X \right) \) the matrix defined by \(\left[ \sigma_{ij}(X) \right]_{kl} = X_{\sigma_{ij}(k)\sigma_{ij}(l)}\), that is, \(\sigma_{ij}(X)\) is the matrix obtained from \(X\) by swapping the \(i\)th and \(j\)th rows and columns. 

\begin{definition}[Symmetry]
    Let \(t^*\) be a positive real number. We say that two players \(i, j \in N\) are \emph{symmetric} during \([0, t^*]\) if we have \(\sigma_{ij} \left( A^{(t)} \right) = A^{(t)}\) at all time \(t \in [0, t^*]\). 
\end{definition}

Suppose players \(i\) and \(j\) are symmetric with respect to \(A\). This means they interact with the network in the same way: they receive the same benefit from every other player \(k \in N \setminus \{i,j\}\), i.e., \(A_{ik} = A_{jk}\), and, in turn, every other player benefits equally from them, meaning \(A_{ki} = A_{kj}\). Additionally, \(i\) and \(j\) benefit equally from each other, so \(A_{ij} = A_{ji}\), and consequently, they derive the same benefit from themselves, implying \(A_{ii} = A_{jj}\). This is visualised in the example of Figure~\ref{fig:indepsymm}, where player \(a\) and \(b\) are symmetric players. 

\begin{proposition}\label{prop: symmetry}
    Two symmetric players \(i\) and \(j\) during \([0, t^*]\) carry the same indirect responsibility on the subnetwork spanned by \(N \setminus \{i, j\}\) on the time interval \([0, t^*]\). 
\end{proposition}

\begin{proof}
    Let \(t^*\) be a positive real number, let \(i \in N\) be an independent player during \([0, t^*]\) and \(t \in [0, t^*]\) be arbitrary. First, notice that 
    \[
    \sigma_{ij} \left( \calL^{(t)} \right) = \sigma_{ij} \left( I - A^{(t)} \right) = \sigma_{ij} \left( I \right) - \sigma_{ij} \left( A^{(t)} \right) = I - A^{(t)} = \calL^{(t)}. 
    \]
    Let us denote by \(\nu_{i | j}\) the indirect responsibility of \(i\) coming from everyone else except \(j\). Then, the influence of the player \(i\) on the subnetwork \(N \setminus \{i, j\}\) is given by 
    \[
    \partial_t \nu_{i | j} (t) = - \sum_{k \in N \setminus \{i, j\}} \calL_{ik}^{(t)} \rho_k[t] = - \sum_{k \in N \setminus \{i, j\}} \calL_{jk}^{(t)} \rho_k[t] = \partial_t \nu_{j | i} (t). 
    \]
    Because \(\nu_{i|j}(0) = 0 = \nu_{j|i}(0)\), the proof is complete. 
\end{proof}

The key idea here is to say that two symmetric players, aside from their direct interaction, exert the same influence on the rest of the network. This is related to the concept of \emph{modules} in unweighted graphs or digraphs~\cite{mcconnell2005linear}. Modules can be understood as sets of nodes that cannot be distinguished by nodes outside of the modules, based solely on their edges. However, nodes within the module may have different connections with other nodes inside the module. A node outside a module is either connected to all members of the module, or to none. Hence, if we permute the nodes inside the module, the nodes outside the module do not notice it.

\medskip

\begin{example}
Consider the network shown in Figure~\ref{fig:indepsymm}. Let the initial responsibility vector be \(f = (0, 0, 0, 5)^{\top}\), meaning only player \(d\) starts with nonzero responsibility. Additionally, player \(c\) continuously generates new impacts according to the function
\(s_c(t) = 0.4t + 0.5\cos(2t)\) 
which is always non-negative but can be decreasing. 
In this network, player \(d\) is an independent player, while players \(a\) and \(b\) are symmetric. Figure~\ref{fig:respindepsymm} displays the responsibility that is allocated to each player, alongside the source function of player \(c\) and the curve
\(5 + 0.2t^2 + 0.25\sin(2t)\),
which represents the total initial responsibility plus the integral of the source function.
We observe that this curve exactly matches the sum of all individual responsibilities, thereby confirming Proposition~\ref{prop:efficiency}. Furthermore, player \(d\)’s responsibility, initially nonzero, declines to zero over time: as an independent player, she neither benefits from others nor from herself, while other players do benefit from her. Finally, players \(a\) and \(b\) receive identical levels of responsibility throughout, which is in line with them being symmetric. 

    \begin{figure}
        \centering
        \includegraphics[width=0.8\linewidth]{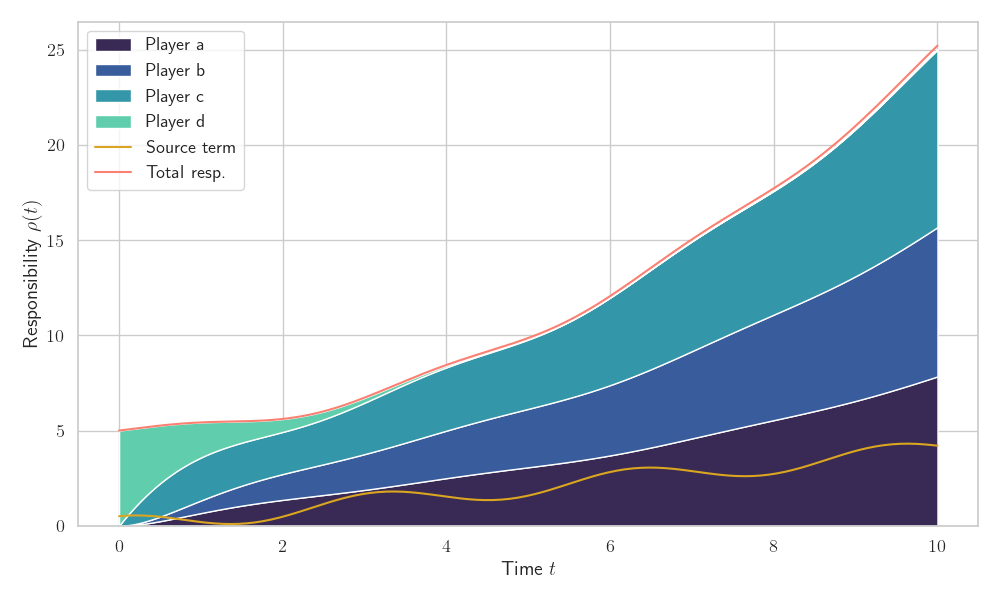}
        \caption{Responsibility allocation in the network of Figure~\ref{fig:indepsymm}, with \(f = (0,0,0,5)^{\top}\), and \(s_c(t)=  0.4t + 0.5\cos(2t)\).}
        \label{fig:respindepsymm}
    \end{figure}
\end{example}

\medskip 

To summarize, our responsibility value \(\rho\) satisfies efficiency, linearity, symmetry and the independent player property. Linearity and efficiency are two natural and desirable properties when allocating responsibility among players. Efficiency ensures that there is no double counting, no loss, and no artificial creation of impact: what is allocated is equal to what is generated. Linearity, on the other hand, prevents strategic manipulation: players cannot reduce their responsibility by splitting, merging, or bundling their emissions differently. If we decompose responsibility into a direct and an indirect component, it is reasonable to expect that direct responsibility should only decrease when a player derives no benefit from the network. For indirect responsibility, we expect players in identical positions within the network to receive identical allocations. 
%These four properties are intuitive, and in fact, they closely resemble the axioms used by Shapley to define the Shapley value in cooperative game theory \parencite{shapley1953value}: linearity, efficiency, symmetry and the null-player property. %The Shapley value is computed as the average of all marginal contributions across all possible permutations of players. The marginal contribution represents the amount of value a player adds to the worth of the coalition by joining it. To draw an analogy with the Laplacian matrix, we can write \([\calL^{(t)} \rho]_i  = -\sum_{j\in N}A^{(t)}_{ij}(\rho_i - \rho_j) \). Here, the term \((\rho_j - \rho_i)\) reflects the difference in responsibility between node \(j\) and node \(i\), and the Laplacian aggregates these local differences across all neighbours of \(i\). In this sense, the Laplacian can be interpreted as encoding a kind of local marginal contribution, reflecting how much a player influences, or is influenced by, the responsibility of their neighbours.

\begin{figure}
    \centering
\begin{subfigure}[t]{0.35\textwidth}
\begin{tikzpicture}
    \node[shape=circle,draw=black] (a) at (0,0) {$a$};
    \node[shape=circle,draw=black] (b) at (1, -1) {$b$};
    \node[shape=circle,draw=black] (c) at (2,-2) {$c$};
     \node[shape=circle,draw=black] (d) at (3,-3) {$d$};
    \path [->] (b) edge[left]  node {\(1.0\)} (a);
    \path [->] (c) edge[left]  node {\(1.0\)} (b);
    \path [->] (d) edge[left]  node {\(1.0\)} (c);
    \path [->] (d) edge[loop below]  node {\(1.0\)} (d);
\end{tikzpicture}
\end{subfigure}
\hfill
\begin{subfigure}[t]{0.6\textwidth}
\includegraphics[width=\textwidth]{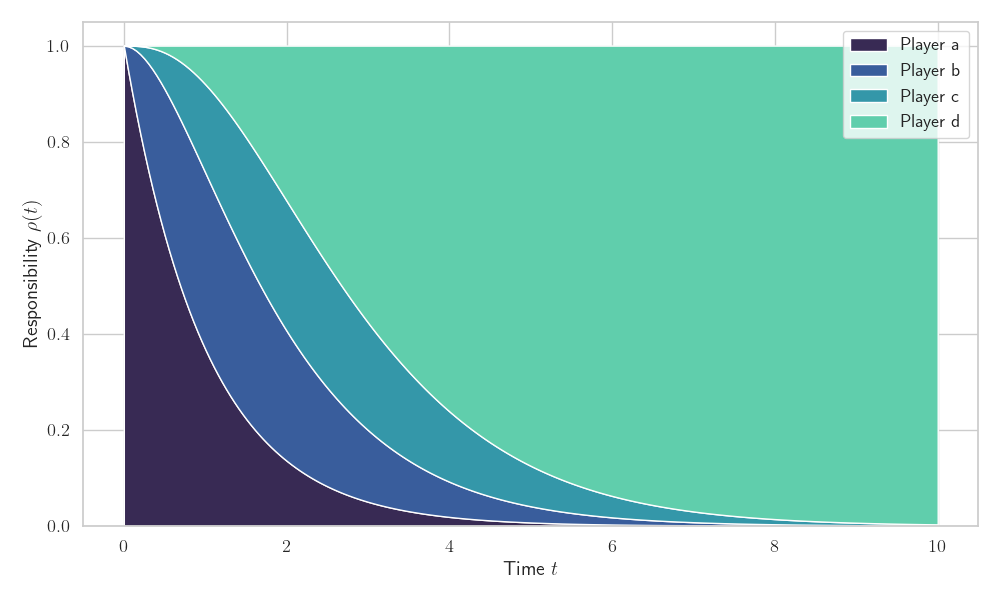}
\end{subfigure}
\vspace{4em}

\begin{subfigure}[t]{0.35\textwidth}
\begin{tikzpicture}
    \node[shape=circle,draw=black] (a) at (0,0) {$a$};
    \node[shape=circle,draw=black] (b) at (-1, 1) {$b$};
    \node[shape=circle,draw=black] (c) at (1,1) {$c$};
     \node[shape=circle,draw=black] (d) at (0,-1.5) {$d$};
    \path [->] (b) edge[left]  node {\(\frac{1}{3}\)} (a);
    \path [->] (c) edge[left]  node {\(\frac{1}{3}\)} (a);
    \path [->] (d) edge[left]  node {\(\frac{1}{3}\)} (a);
    \path [->] (d) edge[loop below]  node {\(1.0\)} (d);
     \path [->] (b) edge[loop left]  node {\(1.0\)} (d);
      \path [->] (c) edge[loop right]  node {\(1.0\)} (d);
\end{tikzpicture}
\end{subfigure}
\hfill
\begin{subfigure}[t]{0.6\textwidth}
\includegraphics[width=\textwidth]{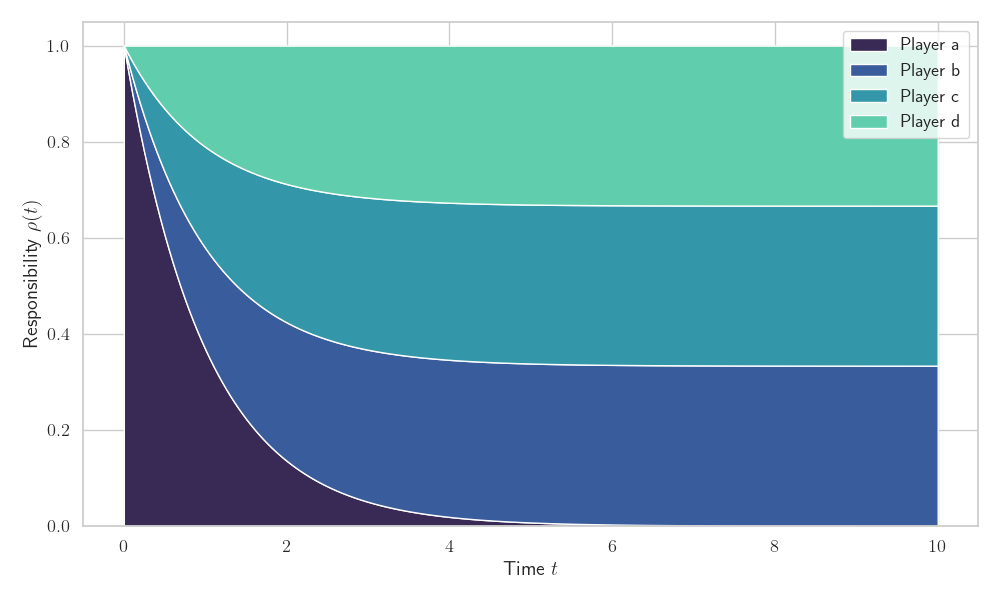}
\end{subfigure}
\caption{A chain network (top) and a star network (bottom) and their corresponding responsibility allocations \(\rho[t]\). Player \(a\) carries an initial responsibility of \(1\), while all other initial responsibilities are zero. We assume the network to be static and we let the source function be zero for everyone. }
\label{fig:chainstar}
\end{figure}

\begin{example}
In this example, we have a set of four players, denoted by \({a,b,c,d}\). We assign player \(a\) an initial responsibility of \(1\), with all other players starting at zero. The source function is assumed to be zero for all players, and the network remains static throughout.
We begin with a chain network, depicted at the top of Figure~\ref{fig:chainstar}. In this case, each player has only one player benefitting from her impacts, and accordingly, all the nonzero edges have weight 1. In this setting, we observe that the initial responsibility of player \(a\) gradually transfers downstream, ultimately concentrating entirely on player \(d\). The total responsibility remains constant over time, consistent with the fact that there are no new impacts being generated. 
We next examine a star network, shown at the bottom of Figure~\ref{fig:chainstar}, with player \(a\) at the center. In this configuration, all other players (\(b, c, d\)) benefit equally from player \(a\), and in addition, each of them fully benefits from themselves. Here, we find that player \(a\)’s responsibility again decays to zero over time, but unlike the chain network, the final responsibility is distributed equally among players \(b, c,\) and \(d\). 

%This behaviour is consistent with Proposition~\ref{prop:eigenvector}, as the responsibility vectors \((0, 0, 0, 1)^{\top}\) for the chain and \((0, \tfrac{1}{3}, \tfrac{1}{3}, \tfrac{1}{3})^{\top}\) for the star are eigenvectors of the respective adjacency matrices. Moreover, consistent with Corollary~\ref{cor: independent-player}, we see that the star network converges significantly faster to its steady state than the chain network.
\end{example}

\section{Approximate solution}\label{sec:approx}
As discussed in Section~\ref{sec:allocating}, when the network changes over time, Equation~\eqref{eq: theorem} does not, in general, admit a closed-form solution. This is due to the non-commutativity of the Laplacian at different times, which prevents the use of standard exponential expressions. Instead, the solution is described using the Peano–Baker series, which constructs a time-dependent state transition matrix that correctly orders the Laplacians over time. While this approach is mathematically rigorous, it still requires numerical approximation in practice. Moreover, the data available in most applications, such as sectoral input-output data or annual greenhouse gas emissions, is inherently discrete. Thus, approximation is not only necessary but also natural for practical use. Although developing such closed-form approximation is beyond the scope of this paper, and has been addressed in prior work for example by \textcite{dacunha2005transition}, we will show in this section that approximating and discretizing our responsibility allocation does not compromise its properties.

\medskip 

The goal is to get a time discretization using the \emph{(forward) Euler method}, by using a difference equation that approximates the differential equation defining the responsibility value, without using differential operators. Let \(h\) be a small positive number. Using Taylor expansions, we have that 
\[
\rho[t_0 + h] = \rho[t_0] + h \partial_t \rho[t_0] + O(h^2).
\]
Hence, the solution to the equation 
\[
\tilde{\rho}[t_0 + h] - \tilde{\rho}[t_0] = h \partial_t \tilde{\rho}[t_0],
\]
is an approximation of the responsibility value that has an error that is proportional to the square of the time step. We can replace the derivative of \(\tilde{\rho}\) using the equation defining \(\rho\) to get 
\[
\tilde{\rho}[t_0 + h] - \tilde{\rho}[t_0] = - h \mathcal{L}^{(t_0)} \rho[t_0] + hs(t_0).
\]
If the value of \(\tilde{\rho}[t_0]\) is known, then we recursively compute \(\tilde{\rho}[t_0 + h]\), then \(\tilde{\rho}[t_0 + 2h]\) and so on, by iteratively applying the update rule. From now on, we fully embrace the discrete nature of time and denote \(\tilde{\rho}_k = \tilde{\rho}[t_0 + kh]\), and set \(\tilde{\rho}[t_0] = \rho[t_0] = f\). This defines a sequence
\begin{equation}\label{eq: discretization}
\tilde{\rho}_{k+1} = \tilde{\rho}_k - h \mathcal{L}^{(k)} \tilde{\rho}_k + h s_k, \quad \text{ initialized by } \quad \tilde{\rho}_0 = f,
\end{equation}
with \(\mathcal{L}^{(k)} = \mathcal{L}^{(t_0 + kh)}\) and \(s_k = s(t_0 + kh)\). Equation~\eqref{eq: discretization} therefore defines our approximation for the responsibility value. The inductive part of the definition of the sequence can be rewritten as 

\begin{equation}
\tilde{\rho}_{k+1} = (1-h)\tilde{\rho}_k + h (A^{(k)} \tilde{\rho}_k + s_k),
\label{eq:convexh}
\end{equation}
which can be interpreted as follows: the new value at the next time step is a combination of the previous value and a transformed version of it, resembling a Markov process with additional impacts. This combination is convex when the \(h\) is sufficiently small. 

\medskip 

In a similar way, applying the forward Euler method to the equations of direct and indirect responsibilities yields
\[
\tilde{\nu}_i^{(k+1)} = \tilde{\nu}_i^{(k)} - h \sum_{j \in N \setminus \{i\}} \mathcal{L}_{ij}^{(k)} \tilde{\rho}_j^{(k)}, \qquad \text{and} \qquad \tilde{\mu}_i^{(k+1)} = \tilde{\mu}_i^{(k)} - h \mathcal{L}_{ii}^{(k)} \tilde{\rho}_i^{(k)} + h s_i^{(k)},
\]
initialized at \(\tilde{\nu}_i^{(0)} = 0\) and \(\tilde{\mu}_i^{(0)} = f\).

\medskip 

In order to be a valuable tool for practical use, we now prove that \(\tilde{\rho}\) still satisfies the properties of \(\rho\) discussed in Section~\ref{sec:prop}. To avoid confusion about the initial condition and the impact sequence, we write \(\tilde{\rho}\left(f, s\right)\) to denote the sequence defined by Equation~\eqref{eq: discretization}, with \(f \in \mathbb{R}_+^n\) and \(s \in \left(\mathbb{R}_+^n\right)^{\mathbb{N}}\). We start with efficiency.

\begin{proposition}
    The approximation is efficient. 
\end{proposition}

\begin{proof} 
We prove it by induction on \(k\). First, we have that \(\tilde{\rho}_0 = f\), hence \(\lVert \tilde{\rho}_0 \rVert_1 = \lVert f \rVert_1\). Now, assume that there exists \(k \in \mathbb{N}\) such that \(\tilde{\rho}_k\) is efficient, i.e., 
\begin{equation}\label{eq: discrete-efficiency}
\lVert \tilde{\rho}_k \rVert = \lVert f \rVert_1 + h  \sum_{l = 0}^k \lVert s_l \rVert_1. 
\end{equation}
Recall that \(f\) and \(s_k\) are all non-negative, for all \(k \in \mathbb{N}\). Hence, we have 
\[
\lVert \tilde{\rho}_{k + 1} \rVert_1 = \mathbf{1}^\top \tilde{\rho}_{k + 1} = \mathbf{1}^\top \left( \tilde{\rho}_k - h \mathcal{L}^{(k)} \tilde{\rho}_k + h s_k \right)
\]
Using that the columns of the Laplacian sum to zero, we obtain 
\[
\begin{aligned}
\lVert \tilde{\rho}_{k + 1} \rVert_1 &= \mathbf{1}^\top \tilde{\rho}_k + h\mathbf{1}^\top s_k = \mathbf{1}^\top f + h \sum_{l = 0}^k \mathbf{1}^\top s_l = \lVert f \rVert_1 + h \sum_{l = 0}^k \left\lVert s_l \right\rVert_1,
\end{aligned}
\]
which is the discrete equivalent of the efficiency property. 
\end{proof}

Let us have a closer look at Equation~\eqref{eq: discrete-efficiency}. We can interpret \(h\) as the width of some rectangles we use to approximate the integral of \(s\), that we mutiply by the norms of \(s\). This resembles a finite approximation of the Riemann-like integral of \(s\). Indeed, each \(s_l\) repesents some discrete values of the function separated with an interval of length \(h\). If \(h\) tends to zero, as the error is quadratic in \(h\),  we should get an error of zero. If we aim for \(\tilde{\rho}_k\) to be an approximation of \(\rho[t]\), we need \(kh\) to be the same order of magnitude as \(t\). If \(h\) gets smaller, the sum will have more terms which means \(k\) should increase. This explains why the term \(h\) is still existent in the definition of efficiency, and, in Equation~\eqref{eq: discrete-efficiency}, the small \(h\) compensates the increase of terms in the sum due to a large \(k\). 

\medskip 

The approximation thus preserves the fact that the total allocated responsibility is equal to the total impacts generated. As a result, there is no double counting or artificial creation or loss of responsibility. We continue with linearity.

\begin{proposition}
    The approximation is linear with respect to the pair \((f, s)\). 
\end{proposition}

\begin{proof}
    Let \(\overline{f}, \hat{f} \in \mathbb{R}_+^n\) and \(\overline{s}, \hat{s} \in \left( \mathbb{R}_+^n \right)^{\mathbb{N}}\). We write \(\overline{\rho}\) for \(\tilde{\rho}(\overline{f}, \overline{s})\) and \(\hat{\rho}\) for \(\tilde{\rho}(\hat{f}, \hat{s})\). We need to show that \(\overline{\rho} + \hat{\rho}\) is a solution of
    \[
    \tilde{\rho}^{(k+1)} = \tilde{\rho}^{(k)} - h \mathcal{L}^{(k)} \tilde{\rho}^{(k)} + h (\overline{s} + \hat{s})^{(k)}, \qquad \text{initialized by} \qquad \tilde{\rho}^{(0)} = \overline{f} + \hat{f}. 
    \]
    Yet, we have
    \[ \begin{aligned} 
    \left( \overline{\rho} + \hat{\rho} \right)^{(k + 1)} = \overline{\rho}^{(k + 1)} + \hat{\rho}^{(k + 1)} & = \overline{\rho}^{(k)} - h \mathcal{L}^{(k)} \overline{\rho}^{(k)} + h \overline{s}^{(k)} + \hat{\rho}^{(k)} - h \mathcal{L}^{(k)} \hat{\rho}^{(k)} + h \hat{s}^{(k)} \\
    & = \left( \overline{\rho}^{(k)} + \hat{\rho}^{(k)} \right) - h \mathcal{L}^{(k)} \left( \overline{\rho}^{(k)} + \hat{\rho}^{(k)} \right) + h \left( \overline{s}^{(k)} + \hat{s}^{(k)} \right) \\
    & = \left( \overline{\rho} + \hat{\rho} \right)^{(k)} - h \mathcal{L}^{(k)} \left( \overline{\rho} + \hat{\rho} \right)^{(k)} + h \left( \overline{s} + \hat{s} \right)^{(k)}. 
    \end{aligned} \]
    Moreover, we notice that \(\left( \overline{\rho} + \hat{\rho} \right)^{(0)} = \overline{\rho}^{(0)} + \hat{\rho}^{(0)} = \overline{f} + \hat{f}\). Hence, \(\overline{\rho} + \hat{\rho}\) is solution, and the approximation is indeed linear with respect to \((f, s)\). 
\end{proof}

So, when approximating the responsibility, it is still guaranteed that splitting or merging of players does not affect the allocation, and the outcome does not depend on whether values are aggregated before or after applying the allocation. Next, we evaluate the responsibility of players that do not benefit from anyone else.

\begin{proposition}\label{prop: approx-independent}
    The approximate indirect responsibility of an independent player is zero. 
\end{proposition}

\begin{proof}
The approximation of the indirect responsibility of an independent player \(i \in N\) is given by 
\[
\tilde{\nu}_i^{(k+1)} = \tilde{\nu}_i^{(k)} - h \sum_{j \in N \setminus \{i\}} \mathcal{L}_{ij}^{(k)} \tilde{\rho}_j^{(k)}, \quad \text{ initialized by } \quad \tilde{\nu}_i^{(0)} = 0.
\]
Because \(i\) is independent, we have that \(A_{ij}^{(k)} = 0\) for all \(j \in N\), and for all \(k \in \mathbb{N}\). Then, for \(j \neq i\), we have \(\mathcal{L}^{(k)}_{ij} = 0\) as well. Hence, the evolution of the approximation of the indirect responsibility of an independent player is given by 
\[
\tilde{\nu}_i^{(k+1)} = \tilde{\nu}_i^{(k)}.  \]
Because it is initialized at \(0\), it is \(0\) for all \(k \in \mathbb{N}\). 
\end{proof}

\begin{corollary}
    The approximate indirect responsibility of an independent player is given by 
    \[
    \tilde{\mu}_i^{(k)} = (1-h)^k f_i + h \sum_{l=0}^{k-1} (1-h)^{k-l-1} s_i^{(l)}. 
    \]
\end{corollary}

\begin{proof}
    By Proposition~\ref{prop: approx-independent}, we have, for all \(k \in \mathbb{N}\), that \(\tilde{\mu}_i^{(k)} = \tilde{\rho}_i^{(k)}\). Moreover, because \(A_{ij}^{(k)} = 0\) for all \(j \in N\) and \(k \in \mathbb{N}\), we have that \(\mathcal{L}_{ii}^{(k)} = 1\). So, we have to find a closed-form expression of \(\tilde{\mu}_i^{(k)}\) from the recursive expression 
    \begin{equation}\label{eq: approx-convex}
    \tilde{\mu}_i^{(k)} = \tilde{\mu}_i^{(k-1)} - h \tilde{\mu}_i^{(k-1)} + h s_i^{(k-1)} = (1-h) \tilde{\mu}_i^{(k-1)} + h s_i^{(k-1)}, \qquad \text{with} \qquad \tilde{\mu}_i^{(0)} = f_i. 
    \end{equation}
    Plugging Equation~\eqref{eq:convexh} into itself recursively gives 
    \[ \begin{aligned} 
    \tilde{\mu}_i^{(k)} & = (1-h)^2 \tilde{\mu}_i^{(k-2)} + (1-h) h s_i^{(k-2)} + h s_i^{(k-1)} \\ 
    & = (1-h)^3 \tilde{\mu}_i^{(k-3)} + (1-h)^2 h s_i^{(k-3)} + (1-h)h s_i^{(k-2)} + h s_i^{(k-1)} \\
    & = \hspace{2cm} \vdots \hspace{2cm} \vdots \hspace{2cm} \vdots \\
    & = (1-h)^k \tilde{\mu}_i^{(0)} + h \sum_{l = 0}^{k-1} (1-h)^{k-l-1} s_i^{(l)}. 
    \end{aligned} \]
    Replacing \(\tilde{\mu}_i^{(0)}\) by \(f_i\) finishes the proof. 
\end{proof}

Notice that, when \(h\) is sufficiently small, the approximation of the direct responsibility of an independent player at a specific step given by Equation~\eqref{eq:convexh} is a convex combination between its previous indirect responsibility and its impact. Lastly, we verify that the symmetry property is preserved in the approximation.

\begin{proposition}
    Two symmetric players \(i\) and \(j\) during \([0, t^*]\) carry the same indirect approximate responsibility on the subnetwork spanned by \(N \setminus \{i, j\}\) on the time interval \([0, t^*]\). 
\end{proposition}

\begin{proof}
Recall that in the proof of Proposition~\ref{prop: symmetry}, we have shown that \(\sigma_{ij} \left( \mathcal{L} \right) = \mathcal{L}\) at all time \(t \in [0, t^*]\) and for all pairs of symmetric players \(i\) and \(j\). The approximation of the indirect responsibility is given by 
\[
\tilde{\nu}_{i|j}^{(k + 1)} = \tilde{\nu}_{i|j}^{(k)} - h\sum_{l \in N \setminus \{i, j\}} \mathcal{L}^{(k)}_{il} \tilde{\rho}_l^{(k)}. 
\]
Using the symmetry property of \(\mathcal{L}^{(k)}\) for all natural \(k\) gives
\[
\tilde{\nu}_{i|j}^{(k+1)} - \tilde{\nu}_{i|j}^{(k)} = - h \sum_{l \in N \setminus \{i, j\}} \mathcal{L}^{(k)}_{il} \tilde{\rho}_l^{(k)} = - h \sum_{l \in N \setminus \{i, j\}} \mathcal{L}_{jl}^{(k)} \tilde{\rho}_l^{(k)} = \tilde{\nu}_{j|i}^{(k+1)} - \tilde{\nu}_{j|i}^{(k)}.
\]
Together with the same initialization \(\tilde{\nu}_{i|j}^{(0)} = \tilde{\nu}_{j|i}^{(0)} = 0\), these define the same sequences, so the indirect responsibilities of \(i\) and \(j\) are identical.
\end{proof}

To summarize, when approximating the solution for the allocation of responsibility, we find that all four properties of efficiency, linearity, the independent player property and the symmetry property are preserved. This demonstrates that the responsibility allocation we propose is not only theoretically robust but also suitable for practical implementation.

\section{Discussion and outlook}
\label{sec:discussion}

In this paper, we have developed a model to describe how responsibility is embedded in a network and how it diffuses from one player to another, based on the characteristics of the underlying network. The players in the network produce some negative impact, as a by-product of their economic activities, which could for example be the emissions of greenhouse gases, the loss of biodiversity, or some (climate-)risk. We have derived a differential equation whose solution satisfies a set of desirable properties. 
While axiomatic analysis often aims to fully characterize a concept through necessary and sufficient conditions, our approach focused on identifying necessary properties that any responsibility measure must satisfy within the dynamic network framework. Although we do not claim sufficiency, these properties still play an important role: they provide structure, clarify the theoretical foundations of the model, and help distinguish valid responsibility measures from implausible alternatives. As such, this partial axiomatization complements the broader normative motivation and offers a foundation for further refinement. Besides providing a full axiomatization, there are various directions for future research and applications. In this section, we discuss some of these possible applications and future extensions. 

\medskip
Though the examples throughout the paper are mostly based on the intuition of carbon emissions in a supply chain, as we mentioned before, the model could also concern other externalities. For example, systemic risk is a hazard that spreads through a network but of which it is not yet clear how to hold players responsible. Applications of diffusion or contagion models to obtain theoretical results about systemic risk have been proposed by for example \textcite{amini2024limit} who use contagion models to evaluate systemic risk on financial networks,  or \textcite{massai2022equilibria} who evaluate equilibria of systemic risk in saturated networks. We want to emphasize that our model is applicable to \emph{any} kind of network, not just supply chains which tend to have some linear or tree-like shape, but also for example multilayer networks, where each layer represents a distinct type of interaction among the set of agents. Note that if each layer consists of the same set of agents, we call it a multiplex network. An example of such a system can be seen in Figure \ref{fig:multilayer}. Notably, \textcite{poledna2015multi} show that the network associated to systemic risk is a multi-layer network where each layer corresponds to some kind of financial transaction.

\begin{figure}[h]
\centering
\SetCoordinates[xAngle=-10, yAngle =20,yLength=1.2,xLength=0.8,zLength=1.1]
\begin{tikzpicture}[multilayer=3d]
  \Vertex[x =0, y = 0, label=A,layer=1]{A1}
  \Vertex[x=1, y = 1, label=B,layer=1]{B1}
  \Vertex[x=0.5, y =3.2,label = C, layer=1]{C1}
 % \Vertex[x=1.5, y=1.5, label=D,layer=1]{D1}
  \Vertex[x=2,y=3,label=E,layer=1]{E1}
  \Vertex[x=3, y=0.8,label=F,layer=1]{F1}
  \Plane[x=-0.8,y=-0.8,width=4.7,height=4.6, layer = 1, NoBorder]
  \Vertex[x =0, y = 0, label=A,layer=2]{A2}
  \Vertex[x=1, y = 1, label=B,layer=2]{B2}
  \Vertex[x=0.5, y =3.2,label = C, layer=2]{C2}
%  \Vertex[x=1.5, y=1.5, label=D,layer=2]{D2}
  \Vertex[x=2,y=3,label=E,layer=2]{E2}
  \Vertex[x=3, y=0.8,label=F,layer=2]{F2}
  \Plane[x=-0.8, y = -0.8, width=4.7,height=4.6,  layer = 2, NoBorder]
    \Vertex[x =0, y = 0, label=A,layer=3]{A3}
  \Vertex[x=1, y = 1, label=B,layer=3]{B3}
  \Vertex[x=0.5, y =3.2,label = C, layer=3]{C3}
%  \Vertex[x=1.5, y=1.5, label=D,layer=3]{D3}
  \Vertex[x=2,y=3,label=E,layer=3]{E3}
  \Vertex[x=3, y=0.8,label=F,layer=3]{F3}
  \Plane[x=-0.8, y = -0.8, width=4.7,height=4.6,  layer = 3, NoBorder]
  \Edge[](A1)(B1)
  \Edge[](B1)(C1)
  \Edge[](B1)(F1)
  \Edge[](C1)(E1)
 % \Edge[](C1)(D1)
  \Edge[lw=0.5, style=dashed](A1)(A2)
  \Edge[lw=0.5, style=dashed](B1)(B2)
  \Edge[lw=0.5, style=dashed](C1)(C2)
%  \Edge[style=dashed](D1)(D2)
  \Edge[lw = 0.5, style=dashed](E1)(E2)
  \Edge[lw = 0.5, style=dashed](F1)(F2)  
  \Edge[](A2)(B2)
  \Edge[](A2)(C2)
  \Edge[](B2)(E2)
%  \Edge[](D2)(F2)
  \Edge[](E2)(F2)
 % \Edge[](C1)(D1)
  \Edge[lw = 0.5, style=dashed](A2)(A3)
  \Edge[lw=0.5, style=dashed](B2)(B3)
  \Edge[lw=0.5, style=dashed](C2)(C3)
 % \Edge[style=dashed](D2)(D3)
  \Edge[lw = 0.5, style=dashed](E2)(E3)
  \Edge[lw = 0.5, style=dashed](F2)(F3)  
  \Edge[](A3)(B3)
  \Edge[](B3)(F3)
  \Edge[](F3)(C3)
  \Edge[](B3)(E3)
 % \Edge[](C3)(D3)
\end{tikzpicture}
\caption{Example of a multiplex network with three layers.}
\label{fig:multilayer}
\end{figure}

In these settings, responsibility may diffuse both within and across layers. Our framework can easily be extended to account for this by defining a supra-Laplacian operator that governs the combined dynamics across all layers, for example by following the approach of \textcite{tejedor2018diffusion}.

%Following the work of %\textcite{tejedor2018diffusion}, we can define the \emph{supra-Laplacian} for a multi-layer network with \(k\) layers as \(\tilde{\calL}_m = D_m - A_m\), where \(A_m \in \bbR^{Nl \times Nl}\) is the `flattened' adjacency matrix of both the connection within one layer and between layers:  \(A_m = A_{intra} + A_{inter}\) with \(A_{intra} = \bigoplus_{l\in K} A_l\) the direct sum of the layers and \(A_{inter}\) the \(k \times k\) matrix where each element is the \(N \times N\) identity matrix, \(I_N\), thereby obtaining a size of \(Nl \times Nl\). As in the single-layer case, \(D_m\) is the diagonal degree matrix corresponding to \(A_m\). We then obtain
% \[
% \tilde{\calL}_m = 
% \begin{pmatrix*}
% \calL_1 + (k-1)I & - I  & \cdots & - I \\
% -I & \calL_2 + (k-1)I & \cdots & -I \\
% \vdots & \vdots & \vdots & \vdots \\
%  -I &  -I & \cdots & \calL_k + (k-1)I \\
% \end{pmatrix*}. \\
% \]

Such a supra-Laplacian still has the property of it being nonnegative and column-stochastic, we can thus easily apply our formulae and see how responsibility flows not only within a layer, but also between layers. A possible extension could be to refine the diffusion model, for example by embedding a threshold value in the model, following the work of \textcite{acemoglu2015systemic} and \textcite{glasserman2016contagion}.

\medskip 

Another type of network that could be of interest is that of hypergraphs. A hypergraph is a powerful generalization of a traditional graph where the hyperedges allow for multi-vertex instead of just pairwise interactions \parencite{berge1984hypergraphs}.
%Formally, a hypergraph \(H\) is defined as a pair \(H = (N,E)\) where \(N\) is a set of nodes and \(E\) is a set of hyperedges, where each hyperedge is a subset of \(N\), that can contain one or more nodes 
It would be valuable to apply our model to hypergraphs to explore the diffusion of responsibility in such networks. An important step into this direction is to define a Laplacian matrix for a hypergraph. Different methods for this have been proposed, for example by \textcite{louis_2015, chan.etal_2018} who introduce a hypergraph Laplacian operator based on satisfying a Cheeger-type inequality, or by \textcite{chan.liang_2020, pearson.zhang_2015} who formulate a hypergraph Laplacian via a diffusion process respectively with and without mediators. However, applying our responsibility measure to hypergraphs requires deeper analysis, as the dynamics of diffusion must account for higher-order interactions. Unlike in graphs, where the responsibility flows linearly along edges, in hypergraphs the flow within a hyperedge may depend nonlinearly on the full distribution of node values within that edge. This leads to a nonlinear differential equation, which better captures complex group dynamics but also requires more sophisticated analytical tools.%However, these authors show that the Laplacian operator satisfying Cheeger's inequality is necessarily non-linear. \textcite{pearson.zhang_2015} propose a similar method based on a diffusion process. Since different Laplacian formulations are suitable for different applications, future research should focus on identifying the most appropriate definition of the hypergraph Laplacian for modeling responsibility allocation.
\medskip 

In conclusion, we have introduced a method to allocate responsibility for negative impacts within a network. The only required inputs are the direct impacts at each node and the adjacency matrix of the underlying network. One of the key advantages of our model, compared to existing approaches, is its ability to account for dynamic networks and a continuous source of inputs, resulting in a more accurate responsibility measure. Though obtaining precise data on the impact map remains a challenge, financial transaction data or input-output data could be effectively used as the adjacency matrix \(A\) and allow for effective adoptability of the model.

\section*{Acknowledgements}
We wish to thank Mayeul Chavanne and Agnieszka Rusinowska for the helpful discussions we had with them. RvdE has received funding from the European Union’s Horizon 2020 research and innovation programme under the Marie Skłodowska-Curie grant agreement No 956107, “Economic Policy in Complex Environments (EPOC)”. DLM benefited from the support of the FMJH Program PGMO, under project number 2023-0009, and the ANR project HQI-ANR-22-PNCQ-0002. 

%\newpage
\printbibliography

\end{document}